\documentclass[10pt]{article}

\usepackage[english]{babel}

\usepackage{graphics,color}
\usepackage{newlfont,epsfig}

\begin{document}
\date{}
\begin{center}
{\Large\bf Quantum Key Distribution using Continuous-variable non-Gaussian States}
\end{center}
\begin{center}
{\normalsize L.F.M. Borelli, L.S. Aguiar, J.A. Roversi and A. Vidiella-Barranco \footnote{vidiella@ifi.unicamp.br}}
\end{center}
\begin{center}
{\normalsize{ Instituto de F\'\i sica ``Gleb Wataghin'' - Universidade Estadual de Campinas}}\\
{\normalsize{ 13083-859   Campinas  SP  Brazil}}\\
\end{center}
\begin{abstract}
In this work we present a quantum key distribution protocol using continuous-variable non-Gaussian states, homodyne detection
and post-selection. The employed signal states are the Photon Added then Subtracted Coherent States (PASCS) in which one photon is
added and subsequently one photon is subtracted. We analyze the performance of our protocol, compared to a 
coherent state based protocol, for two different attacks that could be carried out by the eavesdropper (Eve).
We calculate the secret key rate transmission in a lossy line for a superior channel (beam-splitter) attack, 
and we show that we may increase the secret key generation rate by using the non-Gaussian PASCS rather than coherent states. 
We also consider the simultaneous quadrature measurement (intercept-resend) attack and we show that the efficiency of Eve's 
attack is substantially reduced if PASCS are used as signal states.
\end{abstract}
\section{Introduction}
The first quantum key distribution (QKD) protocol, conceived in 1984 (BB84) \cite{bb84}, is an inherently discrete protocol; it not only requires 
(discrete) single photon sources, but the modulation of the signals is also discrete. Although the ideal BB84 has been proved unconditionally secure 
\cite{peev09,presk00}, there are still practical shortcomings: reliable single photon sources (used by Alice, the sender) are hard to build, and 
photon counters (used by Bob, the receiver) limit the key generation rate. Notwithstanding fully discrete-variable protocols have been 
successfully accomplished over distances of more than $250$ km in ultra low loss fibres \cite{gisin09}.
Meanwhile, several alternative QKD protocols using other (continuous-variable) light sources have been proposed - employing, for instance,
squeezed states \cite{ralph99,hillery00,cerf01,horak} or coherent states \cite{ralph99,gg02,gg03,borelli}. In such continuous-variable protocols, 
the key may be encoded by Alice in the quadrature variables, and Bob will be allowed to employ photomultipliers (which are faster than single
photon detectors) to read the signals via homodyne detection. 
Continuous-variable protocols may be classified as: i) all continuous protocols \cite{gg02,gg03}, 
for which Alice prepares, for instance, Gaussian states such as coherent states, with random amplitudes drawn from a continuous Gaussian 
distribution, or ii) hybrid protocols \cite{horak,korol,hirano,grangier09}. In the hybrid protocols Alice uses light prepared in continuous-variable 
light signals, but the encoding is made using a discrete set of states (e.g., four states). 
At the same time we are witnessing considerable advances concerning the implementation of QKD in real-world conditions 
\cite{realworld01,realworld02} which usually requires long-distance communication. 
The all continuous-variable protocols are mostly based on coherent states, which are easier to generate than
other quantum states of light. However, coherent state based protocols are normally more effective in shorter ranges, due to poor performance 
in low signal-to-noise ratio conditions. Recently, though, a record of $80$ km has been established for an improved version \cite{grangier13} of the 
GG02 continuous-variable protocol \cite{gg02}. In spite of those advances, it would be interesting to seek other possibilities for long-distance QKD. A 
viable alternative are the hybrid continuous/discrete protocols, which may employ either Gaussian or non-Gaussian states. We would like to remark that 
continuous-variable non-Gaussian states (contrary to Gaussian states) may allow the use of quantum repeaters in order to increase the transmission 
range of a practical QKD system \cite{gran}.

In this paper, we propose a protocol for QKD based on continuous-variable 
non-Gaussian states, viz., photon added then subtracted coherent states 
(PASCS). The PASCS may be generated in a relatively straightforward way departing from a Gaussian 
(coherent) state \cite{bellini07}. We may then formulate a protocol similar to already existing continuous-variable protocols 
\cite{horak,hirano} employing homodyne detection and post-selection \cite{leuchs}. We encode bits $0$ and $1$ in two pairs of PASCS (each pair containing states with opposite phases), which are randomly prepared by Alice. Alice sends light signals to Bob through a lossy line, who will perform homodyne detections on them. 
In order to demonstrate the robustness of our protocol against eavesdropping, we calculate the transmitted secret bit rate, 
$(S_{AB})$ \cite{lutkenhaus2} for a beam splitter attack (superior channel attack), as well as for a kind of intercept-resend
attack (simultaneous measurement quadrature attack). 
That analysis will allow us to assess the security of the protocol using two different attacks as well as to establish a 
comparison with the performance of other protocols. 
Our paper is organized as follows: in Section 2, we briefly introduce the PASCS. 
In Section 3, we review the basic structure of the protocol. 
In Section 4, we analyze the performance of our protocol under the superior channel attack: 
we calculate the secret key rate of our protocol and compare the results with those obtained using a similar protocol using coherent states. In Section 5 we consider a intercept-resend attack: the simultaneous measurement quadrature attack. 
We evaluate the eavesdropper success rate for both the PASCS and the coherent states. 
In Section 6, we discuss the results and present our conclusions.

\section{Photon Added then Subtracted Coherent States}
It is possible to perform quantum state engineering, via conditional measurements by adding and/or subtracting photons of a quantized light field, 
as discussed in \cite{welsch98}. Earlier in the nineties there were envisaged the photon added coherent states (PACS) \cite{agarwal91}, which 
were successfully generated a few years ago \cite{bellini04}. Subsequently, the combination of photon adding and photon subtracting in the 
electromagnetic field has also been experimentally explored \cite{bellini07}. 
In general, the operation of firstly adding $k$ photons and then subtracting $l$ photons from a coherent state $|\alpha\rangle$ 
results in the following state (PASCS) \cite{fan11}:
\begin{equation}
|k,l,\alpha\rangle = [N_{k,l}(\alpha )]^{-1/2} \hat{a}^l\hat{a}^\dagger{}^k |\alpha\rangle,
\end{equation} 
with normalizing constant 
\begin{equation}
N_{k,l}(\alpha )=\sum _{m=0}^l \frac{(l!)^2(l+k-m)!}{(-1)^m m!((l-m)!)^2} L_{l+k-m}\left(-|\alpha |^2\right),
\end{equation}
and where $L_{l+k-m}\left(-|\alpha |^2\right)$ is the Laguerre polynomial of order $(l+k-m)$.

Of particular interest for our purposes, are the PACS and the PASCS having just one photon added and one photon subtracted $(k=l=1)$. 
Thus, from an initial coherent state $|\alpha\rangle$, we first add one photon to it, or $|\phi_A\rangle \propto \hat{a}^\dagger|\alpha\rangle$ 
and then subtract one photon from the resulting state, obtaining the PASCS 
$|1,1,\alpha\rangle\equiv \propto \hat{a}|\phi_A\rangle$. An interesting feature of the state 
$|1,1,\alpha\rangle$ is that it may be written as 
a superposition of a coherent state and a photon added coherent state (PACS), i.e. 
$|1,1,\alpha\rangle \propto \hat{a} \hat{a}^\dagger |\alpha\rangle \propto 
(1 + \hat{a}^\dagger \hat{a})|\alpha\rangle \propto |\alpha\rangle + \alpha |\phi_A\rangle$. 
In other words, this specific PASCS may be written as a superposition of a Gaussian state (coherent state) with a 
non-Gaussian component (PACS) weighted by $\alpha$. 

A useful and well known representation of the field states is the Wigner function - a quasiprobability distribution 
in phase space \cite{wigner32,wigner84}. For a density operator $\hat\rho$, the Wigner 
function may be written as:
\begin{equation}
W(\zeta)=\frac{2}{\pi}\sum_{n=0}^\infty (-1)^n\langle n|\hat{D}^{-1} (\zeta)\hat\rho \hat{D}(\zeta)|n\rangle,
\end{equation}
where $\zeta = \zeta_r + i\zeta_i$, being $(\zeta_r,\zeta_i)$ the phase space coordinates, and $\hat{D}$ is Glauber's displacement operator, 
$\hat{D}(\zeta) = \exp(\zeta\hat{a}^\dagger - \zeta^*\hat{a})$. For the PASCS, $\hat{\rho} = |k,l,\alpha\rangle\langle k,l,\alpha |$, 
the corresponding Wigner function reads \cite{fan11}

\begin{equation}
W^{k,l}(\zeta;\alpha)=\frac{2 e^{(-2|\alpha -\zeta|^2)}}{\pi N_{k,l}(\alpha)}\sum_{n=0}^k \frac{(-1)^n (k!)^2}{n! ((k-n)!)^2}
|H_{k-n,l}(i(2\zeta -\alpha ),i\alpha^*)|^2,\label{wignerfunc}
\end{equation}
being $H$ the bivariate Hermite polynomials
\begin{equation}
H_{p,q}(\epsilon ,\varepsilon )=\sum _{r=0}^{min(p,q)} \frac{(-1)^r p!q!}{r!(p-r)!(q-r)!}\epsilon ^{p-r} \varepsilon ^{q-r}.
\end{equation}

For comparison, we have plotted in Figure 1 the Wigner function of the PASCS 
having just one photon added and one photon subtracted [equation (\ref{wignerfunc}) with $k=l=1$], 
together with the Wigner function of the coherent state $|\alpha\rangle$, given by 
$W(\zeta;\alpha) = \frac{2}{\pi}\exp(-2|\alpha - \zeta|^2)$.
The Wigner function of a coherent state is exactly a Gaussian function, while the PASCS's Wigner function has a slight deformation 
as well as a negative part, a clear indication of the nonclassicality of the state. Apart from being useful for identifying some
features of quantum states, the Wigner function may also be used to analyse the security of our protocol, as we are going
to show below.

\section{The Protocol}

The protocol works as follows: firstly, Alice randomly chooses one of the four PASCS (for $\alpha$ real): 
either $|\psi_{AS+}\rangle \equiv |1,1,\alpha\rangle$ and $|\psi_{AS+i}\rangle\equiv |1,1,i\alpha\rangle$ 
(representing bit $1$), or $|\psi_{AS-}\rangle \equiv |1,1,-\alpha\rangle$ and $|\psi_{AS-i}\rangle\equiv |1,1,-i\alpha\rangle$, 
(representing bit $0$) in the horizontal and vertical bases, respectively. The plots of the Wigner functions corresponding to 
$|\psi_{AS-}\rangle$ and $|\psi_{AS+}\rangle$ in Figure 2 give a clear picture of their distinguishability in phase space.
In a second step Alice sends a light signal prepared in the chosen state to Bob, who randomly selects
either the horizontal or the vertical basis and performs a homodyne detection on the received signal. 
Bob also fixes a convenient value for the post-selection threshold, $\beta_c$.
We now denote $\beta = \beta_r + i\beta_i$ the (complex) measurement variable corresponding to Bob's measurement.
If in a given measurement, he finds $\beta_{r,i} < -\beta_c$, Bob assigns
value 0 the bit; if he finds $\beta_{r,i} > \beta_c$, he assigns value 1 to the bit. Otherwise, Bob tells Alice to neglect 
the corresponding bit. 

\section{Beam-splitter attack: Superior channel attack}

Due to the transmission line losses (imperfect channel), it is possible for an eavesdropper (Eve) to intercept a fraction of the signal 
without being noticed by the legitimate users. To do that, Eve uses an asymmetric beam-splitter of transmissivity $T$ and reflectivity $R$, 
with $T^2 + R^2 =1$. She keeps the reflected part of the beam (the transmitted part is sent to Bob via a lossless channel) stored in a 
quantum memory and waits for the announcement of the measurement basis used by Bob. For simplicity, in this security analysis we consider 
just the case in which the horizontal basis is announced, as the discussion is analogous for the vertical basis due to symmetry.   
To estimate the amount of gain of secret information per transmitted pulse $S_{AB}$ it is necessary to derive an upper 
bound of the information leaked to Eve when she splits the beam, as discussed in \cite{lutkenhaus2,lutkenhaus3}. 
A relevant quantity in the following derivation is the joint measurement probability, of Bob obtaining the result 
$\beta_{r}$ and Eve obtaining $\epsilon_{r}$, 
\begin{equation}
P_{\pm}(\beta_{r},\epsilon_{r})=\int \widetilde{W}_{\pm}(\beta,\epsilon) d\beta_{i}d\epsilon_{i},
\end{equation}
where $\widetilde{W}_{\pm}(\beta,\epsilon)$ is the (two-mode) Wigner function of the beam-splitter output \cite{horak,wu,ou},
\begin{equation}
\widetilde{W}_{\pm}(\beta,\epsilon)= W_{\psi_{AS\pm}}^{1,1}(T \beta - R \epsilon,\alpha)W_{vac}(R\beta+T\epsilon).
\end{equation}
The $\pm$ signs refer to the pair of states we are considering for the security analysis, namely $|\psi_{AS+}\rangle$ 
and $|\psi_{AS-}\rangle$. In the expression above, 
$W_{\psi_{AS\pm}}^{1,1}(T \beta - R \epsilon,\alpha)$ is the (single mode) Wigner function [equation (\ref{wignerfunc})] of the PASCS resulting 
from the addition of only one photon to a coherent state $\alpha$ and subtraction of one photon from the resulting state. In the other port of 
the beam splitter we have the vacuum as input state, with Wigner function $W_{vac}(R\beta+T\epsilon)$.

Because the PASCS is not a coherent state, the two emerging beams from the beam-splitter are normally in an entangled state. 
Thus, the joint probability distribution does not factorize, and the results of 
measurements made by Bob, $\beta_r$ and Eve, $\epsilon_r$ will be somehow correlated, as shown in Figure 3. 
This means that, if Bob measures a relatively large value for his quadrature ($\beta_r$), Eve is likely to measure a small 
value for hers ($\epsilon_r$). For instance, as seen in Figure 3: the maximum of $P_+(\beta_r,\epsilon_r)$ occurs for 
$\beta_r = 1.2$, while $\epsilon_r = -0.70$.
Thus, if we increase the value of the post-selection threshold, the bit error rate on Eve's 
side will also be increased.
 
After performing an ideal error correction and privacy amplification, we may obtain a lower bound for the gain of secret information per 
transmitted pulse, $S_{AB}$ as discussed in \cite{horak,lutkenhaus2,lutkenhaus3,shannon}.
Firstly we define $r_{acc}$, the fraction of accepted bits $r_{acc} = [P(0)+P(1)]/2$, with 
\begin{eqnarray}
P(1)=\int_{\beta_{c}}^{\infty}P'_+(\beta_{r})d\beta_{r} \\
P(0)=\int_{-\infty}^{-\beta_{c}}P'_+(\beta_{r})d\beta_{r},
\end{eqnarray}
and 
\begin{equation}
P'_{\pm}(\beta_{r})=\int \widetilde{W}_{\pm}(\beta,\epsilon) d\beta_{i}d\epsilon_{i}d\epsilon_{r}.
\end{equation}
The Shannon Information $I_{AB}$ is defined as
\begin{eqnarray}
I_{AB} &=& \int_{\beta_c}^\infty d\beta_r \frac{P'_+(\beta_r) + P'_+(-\beta_r)}{P(0)+P(1)} \\ \nonumber
&\times&\left\{1+\delta(\beta_r)\log_2 \delta(\beta_r) +[1-\delta(\beta_r)]\log_2 [1-\delta(\beta_r)]\right\},
\end{eqnarray}
with 
\begin{equation}
\delta(\beta_r) = \frac{P'_+(-\beta_r)}{P'_+(\beta_r)+P'_+(-\beta_r)}.
\end{equation}
The amount of reduction of the raw key during the privacy amplification may be written as $\tau =1+\log_2\left(P_c\right)$,
where $P_{c}$ is the collision probability \cite{horak}
\begin{eqnarray}
P_c=\frac{1}{2}\int \frac{{\cal P}{}^2_+\left(\epsilon _r\left|\beta _c<\right|\beta _r|\right)+
{\cal P}{}^2_-\left(\epsilon _r\left|\beta _c<\right|\beta _r|\right)}{{\cal P}_+\left(\epsilon _r\left|\beta _c<\right|\beta _r|\right)
+{\cal P}_-\left(\epsilon _r\left|\beta _c<\right|\beta _r|\right)} \, d\epsilon _r,
\end{eqnarray}
where
\begin{equation}
{\cal P}_\pm \left(\epsilon _r\left|\beta _c<\right|\beta _r|\right)=\int _{\beta _c<\left|\beta _r\right|}
\frac{P_\pm \left(\beta _r,\epsilon _r\right)}{P(0)+P(1)}d\beta _r
\end{equation}
is Eve's probability distribution conditioned to the fact that a pulse $\pm$ was sent and that Bob accepted the bit in his post-selection. The 
collision probability plays a crucial role in the generation of the secret key, indicating by which amount the raw key must be reduced in order to 
eliminate Eve's knowledge about it. The secret information $S_{AB}$ is thus given by
\begin{equation}
S_{AB}=r_{acc}\left(I_{AB}-\tau \right).
\end{equation}
The results are shown in Figure 4. We have that the maximum of the surface representing the secret information is
$S_{AB}^{max}\approx 0.140$ for the
coherent state while $S_{AB}^{max}\approx 0.167$ for the PASCS. i.e., a percent improvement of about $19\%$
if the PASCS are used in place of coherent states. Moreover, we note that the PASCS based protocol is more efficient for 
smaller values of the amplitude $\alpha$ of the transmitted pulse, compared to the coherent state case, as seen in Figure 4. 
We remind that the PASCS (having just one photon added and one photon subtracted) may be written as a superposition of the coherent state $|
\alpha\rangle$ with a PACS, or $|\psi_{AS}\rangle \propto |\alpha\rangle + \alpha |\phi_A\rangle$; thus, for small $\alpha$ the contribution of the 
PACS (non-Gaussian state) will also be very small, and the PASCS will be close to a coherent (Gaussian) state. Nevertheless, it will still generate an 
entangled state after crossing the beam splitter. This will introduce anti-correlations between Bob's and Eve's measurements results
(see Figure 3), which favours the security of the PASCS-based protocol, given that Bob will be able to reduce Eve's knowledge about the bits via post-
selection. In Figure 5 we have plotted the secret bit rate $S_{AB}$ as a function of transmission distance in a standard optical fibre for protocols
using PASCS and coherent states. 
We note that a PASCS-based protocol outperforms a protocol based solely on coherent states, 
in the sense that a secret key could be generated at higher rates for a given distance.  

\section{Intercept-resend attack: Simultaneous quadrature measurement attack}

For complementarity, we discuss now a second (intercept-resend) attack performed by Eve in which she splits the 
incoming pulses of light in a 50:50 beam-splitter and performs simultaneous quadrature measurements on the outgoing beams. 
She then tries to infer (with probability $P_{corr}$) the state of the signal sent by Alice. Here we consider the 
preparation of four possible states by 
Alice, defined above as $|\psi_{AS\pm(i)}\rangle$. If Eve measures ($\beta_r$, $\varepsilon_i$), she will choose
the state of the signal for which the associated joint probability distribution $P_{\pm(i)}$ is maximum. For each state
we have a corresponding region in phase space (each one of area $A_0$), i.e., $\beta_r \geq |\varepsilon_i|$ for
$|\psi_{AS+}\rangle$; $\varepsilon_i > |\beta_r|$ for $|\psi_{AS+i}\rangle$; $-\beta_r \geq |\varepsilon_i|$ 
for  $|\psi_{AS-}\rangle$ and $-\varepsilon_i > |\beta_r|$ for $|\psi_{AS-i}\rangle$.
Generally speaking, the associated probability distributions are given by
\begin{equation}
P_{\pm(i)}(\beta_{r},\epsilon_{i})=\int \widetilde{W}_{\pm(i)}(\beta,\epsilon) d\beta_{i}d\epsilon_{r},
\end{equation}
where $\widetilde{W}_{\pm}(\beta,\epsilon)$ is the (two-mode) Wigner function of the beam-splitter output,
\begin{equation}
\widetilde{W}_{\pm(i)}(\beta,\epsilon)= W_{\psi_{AS\pm(i)}}^{1,1}(T \beta - R \epsilon,\alpha)* W_{vac}(R\beta+T\epsilon).
\end{equation}
As in reference \cite{horak} we may define Eve's success rate for the attack, $P_{corr}$. For instance, for a signal
in the state $|\psi_{AS+}\rangle$, we have
\begin{equation}
P_{corr}=2 \int_{A_0} P_{+}(\beta_{r},\epsilon_{i})d\beta_{r}d\epsilon_{i}.
\end{equation}
The efficiency of such an attack may then be evaluated. Alice can make an optimization of the coherent amplitude $\alpha$ 
given a fixed error rate $\delta = 1.15\times10^{-3}$ for a lossless line and without the presence of Eve. In Figure 6
we have the optimum $\alpha$ and the fraction of accepted bits, $r_{acc}$ as a function of the post-selection
threshold $\beta_c$, for both coherent state and the PASCS. We note that the value of optimum $\alpha$ (for each $\beta_c$) 
is in general smaller in the PASCS based protocol, compared to the coherent state case. 
Thus, even though the rate of accepted bits
are smaller for the PASCS, the probability of Eve obtaining the correct bit becomes also smaller in this case, given
that the optimum value of $\alpha$ (for a given value of threshold $\beta_c$) is smaller for the PASCS. In Figure 7
we have a plot of $P_{corr}$ as a function of $\beta_c$, which clearly shows the advantage of the PASCS over
coherent states concerning the simultaneous quadrature measurement attack.

\section{Conclusions}
We have shown that a continuous-variable protocol based on PASCS having just one photon added and one photon subtracted
is more efficient than a coherent state-based protocol, both of them using homodyne detection and post-selection as well as
similar amplitudes for the coherent states employed. We have performed a security analysis based on the superior channel attack, 
and concluded that the PASCS-based protocol would allow the legitimate users (Alice and Bob) to build a secret key 
with transmission rates higher than the ones obtained from coherent state based protocols. We have also analyzed the simultaneous 
quadrature measurement attack, and we have shown that Eve's success rate is smaller if PASCS are used in the place of
coherent states. We would like to remark that this work is an attempt to explore the possibilities of utilization
of non-Gaussian states for quantum key distribution purposes, and this may open up new directions for continuous-variable protocols. 
We believe that implementations based on states such as the PASCS could be considered as viable alternatives.

\section*{Acknowledgments}

\noindent This work was partially supported by 
CNPq (Conselho Nacional de Desenvolvimento Cient\'\i fico e Tecnol\'ogico - INCT of Quantum Information), 
FAPESP (Funda\c c\~ao de Amparo \`a Pesquisa do Estado de S\~ao Paulo - CePOF of Optics and Photonics), 
and CAPES (Coordena\c c\~ao de Aperfei\c coamento de Pessoal de Ensino Superior), Brazil.



\begin{figure}[hp]
\centering\includegraphics[trim = 1.5cm 11cm 1.5cm 3cm,clip,width=13cm]{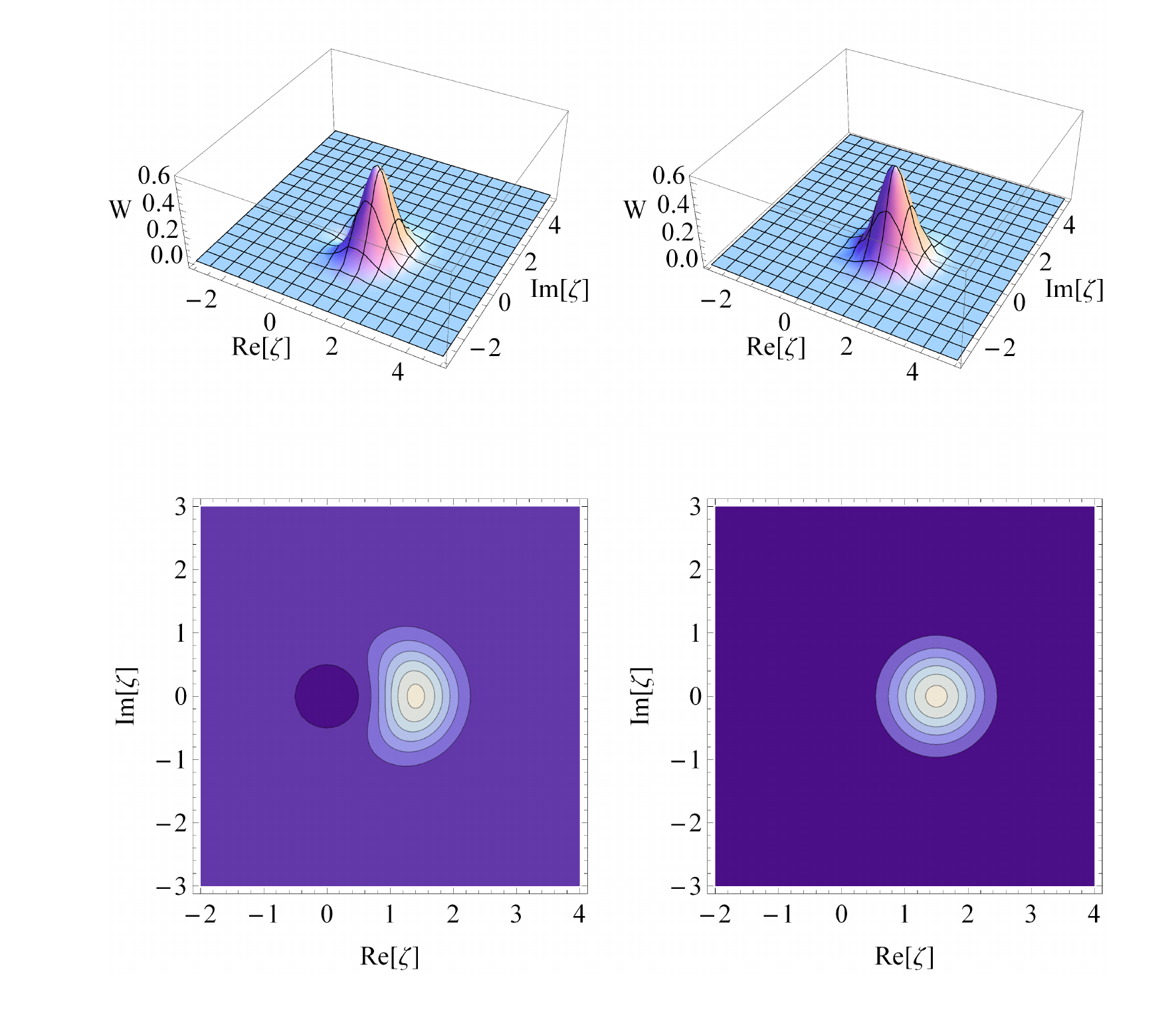}
\caption{Wigner Function and Contour Plots of: PASCS (left) with one photon added and one photon subtracted from a coherent state
having $\alpha = 1$ and a Coherent State (right) having $\alpha' = 1.5$.}
\end{figure}

\begin{figure}[hp]
\centering\includegraphics[trim = 1.5cm 11cm 1.5cm 1cm,clip,width=10cm]{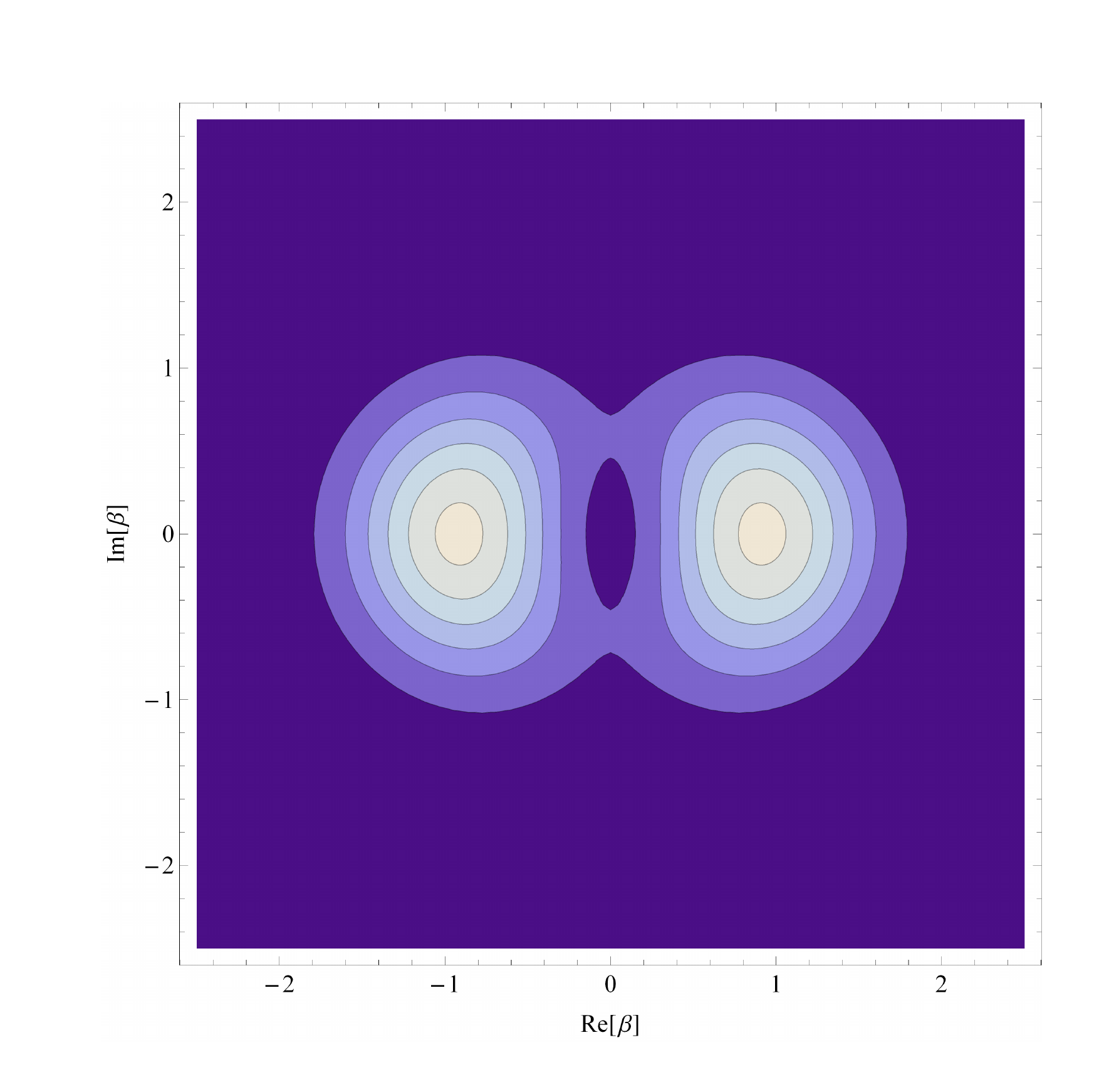}
\caption{Contour plots of the PASCS with one photon added and one photon subtracted from a coherent state
having $\alpha = 0.55$, $(|\psi_{AS+}\rangle)$ and $\alpha = -0.55$ $(|\psi_{AS-}\rangle)$. }
\end{figure}

\begin{figure}[!ht]
\centering\includegraphics[trim = 1.5cm 18cm 3.5cm 2cm,clip,width=13cm]{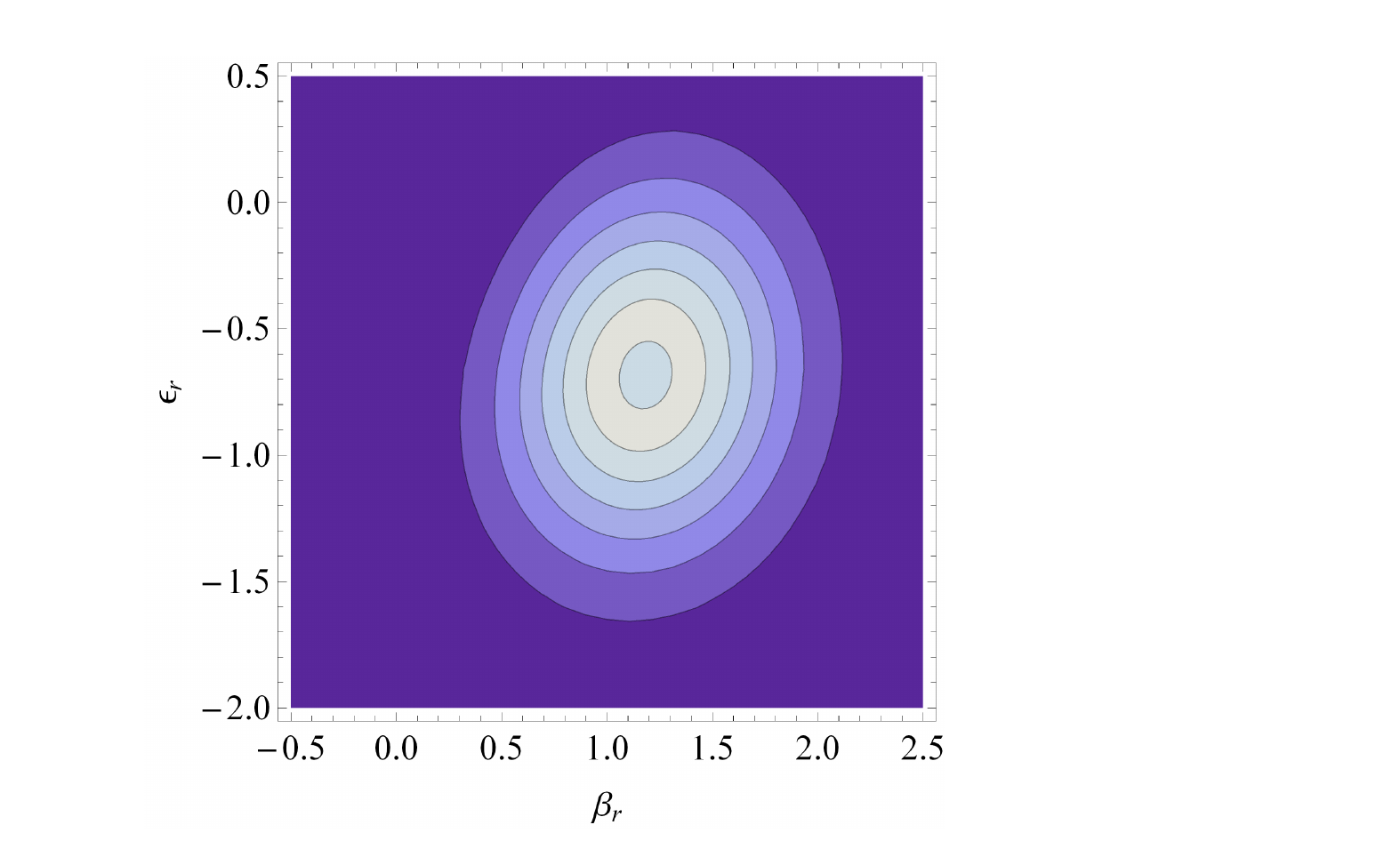}
\caption{Contour plot of the joint probability distribution, $P_{+}(\beta_{r},\epsilon_{r})$, for $\alpha=1$ and $T^2=0.75$.}
\end{figure}

\begin{figure}[!ht]
\centering\includegraphics[trim = 1.5cm 19cm 1.5cm 3cm,clip,width=13cm]{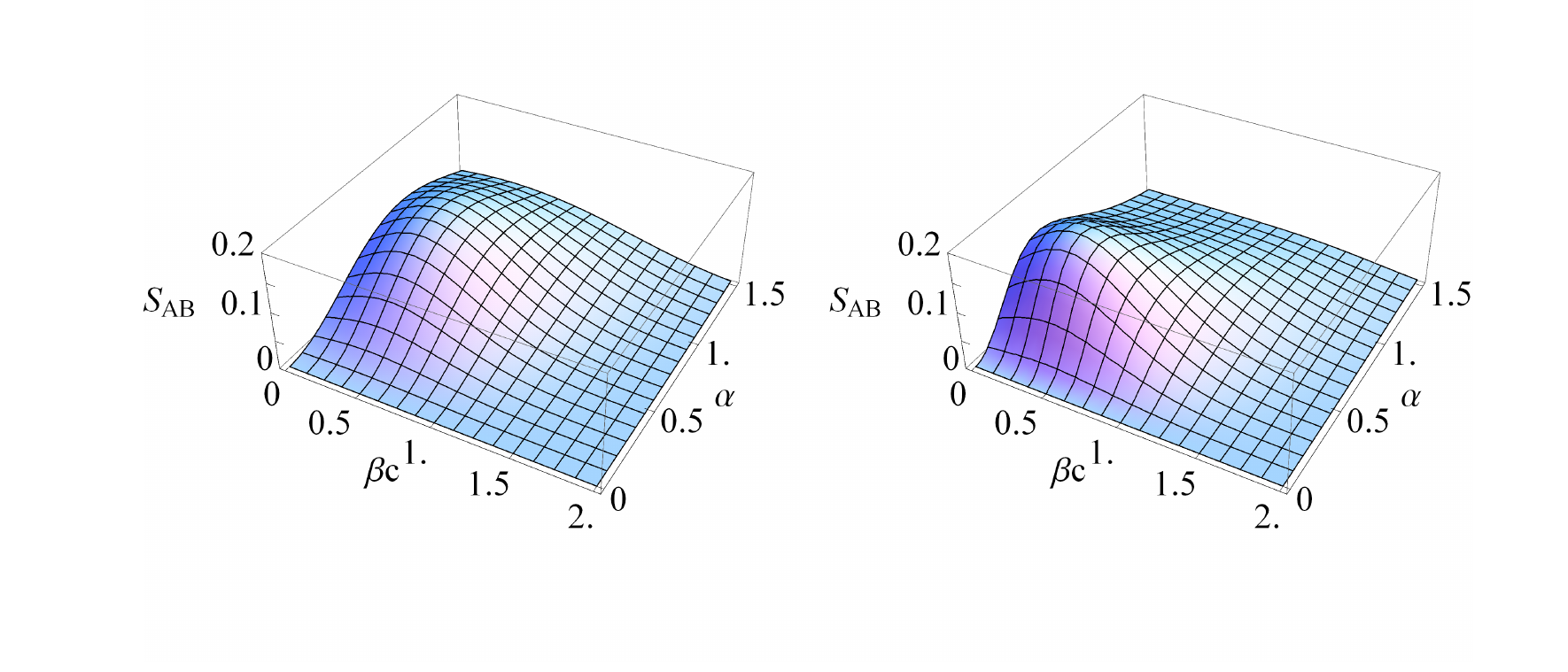}
\caption{Secret key rate $S_{AB}$ versus the coherent state amplitude $(\alpha)$ and the post-selection threshold $(\beta_c)$ 
for a coherent state (left) and for a PASCS having just one photon added and one photon subtracted (right). The channel transmission is $T^2=0.75$.}
\end{figure}

\begin{figure}[!ht]
\centering\includegraphics[trim = 1.5cm 19cm 1.5cm 2cm,clip,width=13cm]{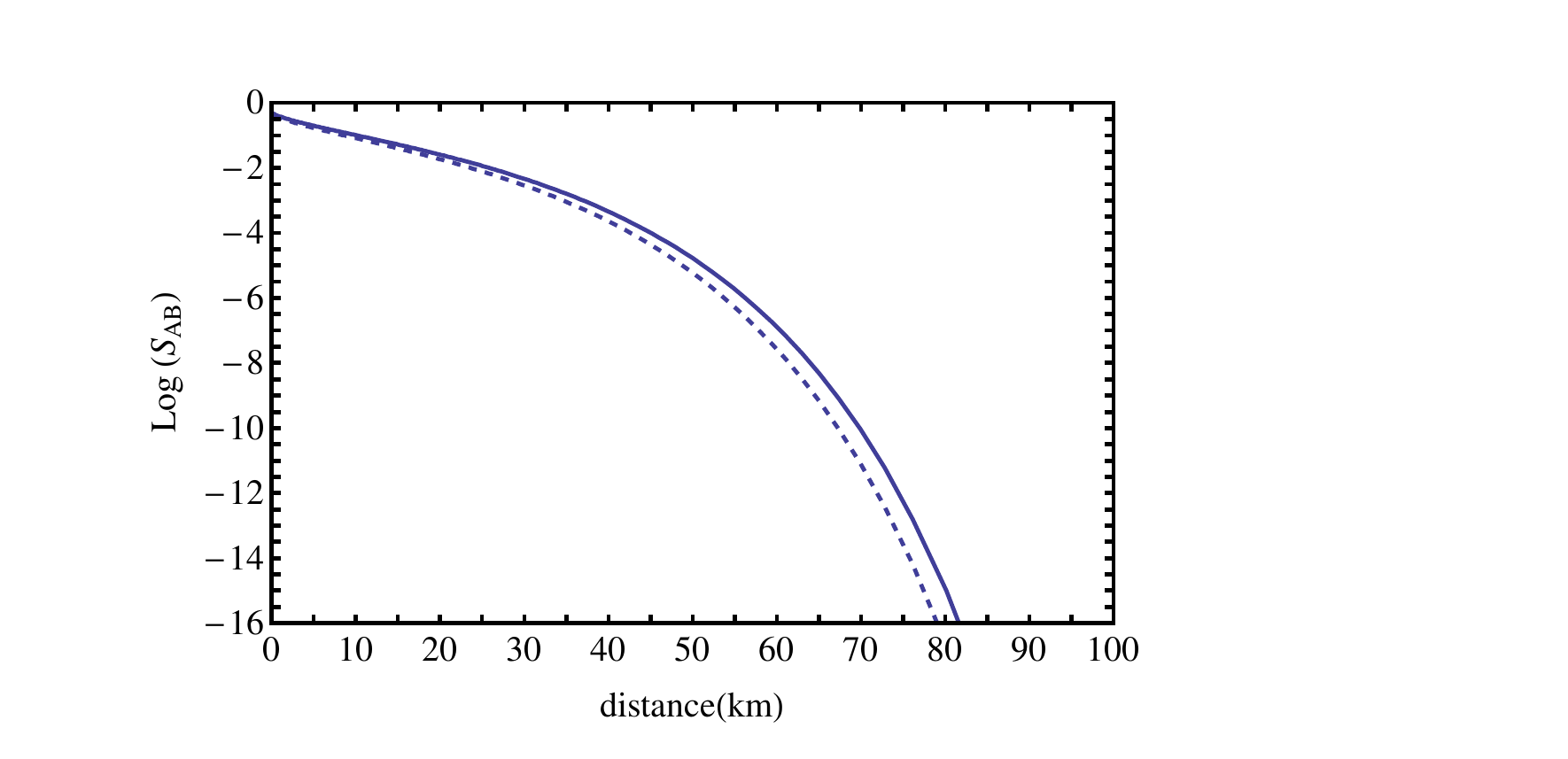}
\caption{Secret key rate $S_{AB}$ versus distance for the: PASCS (solid line) and the 
coherent state (dashed line). We have considered an optical fiber loss coefficient 
$0.2$ dB/km for a wavelength of $1.22\mu$m.}
\end{figure}

\begin{figure}[!ht]
\centering\includegraphics[clip,width=8cm]{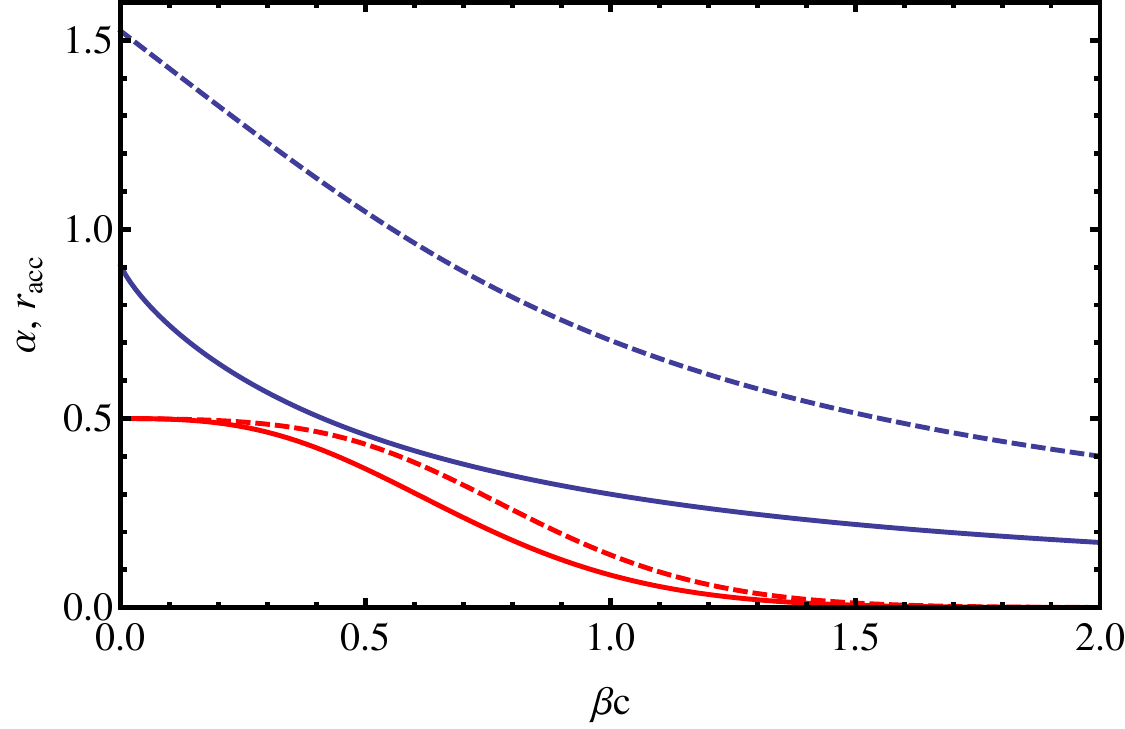}
\caption{Optimum $\alpha$ as a function of $\beta_c$, for the PASCS (solid blue line) and coherent state (dashed blue line);
fraction of accepted bits, $r_{acc}$ as a function of the post-selection
threshold $\beta_c$, for the PASCS (solid red line) and coherent state (dashed red line)}
\end{figure}

\begin{figure}[!ht]
\centering\includegraphics[clip,width=8cm]{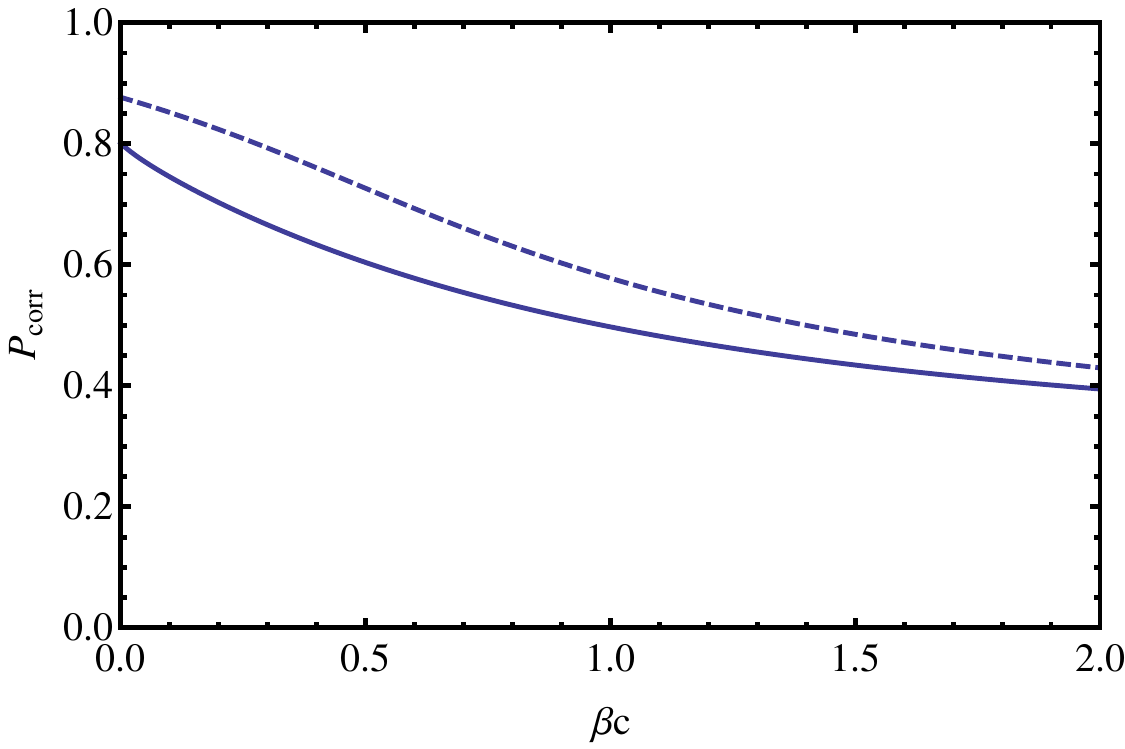}
\caption{The rate of Eve's success $P_{corr}$ as a function of $\beta_c$, PASCS (solid line) and coherent state 
(dashed line).}
\end{figure}

\end{document}